\documentclass{article}
\usepackage[utf8]{inputenc}
\usepackage{amsmath}
\usepackage{amsfonts}
\usepackage{amssymb}
\usepackage{amsthm}
\usepackage{stmaryrd}

\usepackage{tikz}
\usetikzlibrary{patterns}
\newcommand{\point}[1]{\node[circle,inner sep = 0pt,minimum size =0pt]}
\newcommand{\vertex}[1]{\node[circle,inner sep = 0pt,minimum size =3pt,fill = #1]}
\newcommand{\Vertex}[1]{\node[circle,inner sep = 0pt,minimum size =5pt,fill = #1]}

\newcommand{\nat}{\mathbb{N}}

\newtheorem{theorem}{Theorem}[section]
\newtheorem{lemma}[theorem]{Lemma}
\DeclareMathOperator{\cross}{cr}
 
\title{On disjoint paths in acyclic planar graphs}
\author{Guyslain Naves\footnotemark[1]}

\begin{document}






\maketitle
\footnotetext[1]{
Grenoble Universities, France, and McGill University, Montréal, Canada,\\
\phantom{spac}{\tt naves@math.mcgill.ca}
}

\begin{abstract}
We give an algorithm with complexity $O(f(R)^{k^2} k^3 n)$ for the integer multiflow problem on instances $(G,H,r,c)$ with $G$ an acyclic planar digraph and $r+c$ Eulerian. Here, $f$ is a polynomial function, $n = |V(G)|$, $k = |E(H)|$ and $R$ is the maximum request $\max_{h \in E(H)} r(h)$. When $k$ is fixed, this gives a polynomial algorithm for the arc-disjoint paths problem under the same hypothesis.
\end{abstract}

\section{Introduction}

Given a \emph{demand} graph $G$ and a \emph{supply} graph $H$ with $V(H) \subseteq V(G)$, capacities $c : E(G) \to \mathbb{N}$ and requests $r : E(H) \to \mathbb{N}$, the \emph{integer multicommodity flow problem} consists of deciding if there is a multiset $\mathcal{C}$ of cycles in $G+H$ satisfying the following three conditions:
\begin{itemize}
\item[-] for each $C \in \mathcal{C}$, $|C \cap E(H)| = 1$,
\item[-] for each $e \in E(G)$, at most $c(e)$ cycles of $\mathcal{C}$ contain $e$,
\item[-] for each $h \in E(H)$, exactly $r(h)$ cycles of $\mathcal{C}$ contain $h$.
\end{itemize}
This definition holds both for directed and undirected graphs. When the capacity function $c$ is equal to $1$ on every edge (or arc), the problem is called as the \emph{edge-disjoint paths problem}, or the \emph{arc-disjoint paths problem} in the directed case. We call the arcs of $H$ \emph{commodities}. We say that $r+c$ is Eulerian if:
\begin{itemize}
\item[-] $\sum_{e \in \delta^+_G(v)} c(e) + \sum_{e \in \delta^+_H} r(e) - \sum_{e \in \delta^-_G(v)} c(e) - \sum_{e \in \delta^-_H(v)} r(e) = 0$ for each vertex $v$, if $G$ and $H$ are directed,
\item[-] $\sum_{e \in \delta_G(v)} c(e) + \sum_{e \in \delta_H(v)} r(e)$ is even for each vertex $v$, if $G$ and $H$ are undirected.
\end{itemize}

The undirected edge-disjoint paths problem is well-known to be NP-hard even with strong restrictions on the instances, like imposing that $G$ is planar (Kramer and van Leeuwen~\cite{kramer1984cwr}), or that $|E(H)| = 2$ (Even, Itai and Shamir~\cite{eis76}). In~\cite{naves2008mfa}, A. Seb\H{o} and the author gave a survey of the integer multiflow problem with some natural restrictions. It appeared that some combinations of additional constraints define problems that are still open. Among them, the complexity of the integer multiflow problem with $G$ planar, $r+c$ Eulerian and $|E(H)|$ bounded is unknown, in both the directed and undirected cases. In this paper, we address the directed acyclic case.

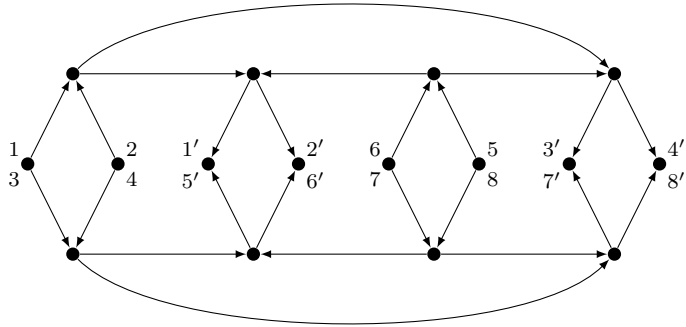
\begin{figure}
\begin{center}
\footnotesize
\begin{tikzpicture}[x=0.3cm,y=0.3cm,>=latex]
\Vertex{black} (nx32y19) at (32,19) {};
\Vertex{black} (nx32y11) at (32,11) {};
\Vertex{black} (nx24y11) at (24,11) {};
\Vertex{black} (nx24y19) at (24,19) {};
\Vertex{black} (nx16y19) at (16,19) {};
\Vertex{black} (nx16y11) at (16,11) {};
\Vertex{black} (nx8y11) at (8,11) {};
\Vertex{black} (nx8y19) at (8,19) {};
\Vertex{black} (nx34y15) at (34,15) {};
\Vertex{black} (nx30y15) at (30,15) {};
\Vertex{black} (nx26y15) at (26,15) {};
\Vertex{black} (nx22y15) at (22,15) {};
\Vertex{black} (nx18y15) at (18,15) {};
\Vertex{black} (nx14y15) at (14,15) {};
\Vertex{black} (nx10y15) at (10,15) {};
\Vertex{black} (nx6y15) at (6,15) {};
\draw[->] (nx8y11) .. controls (12,7) and (28,7) .. (nx32y11);
\draw[->] (nx8y19) .. controls (12,23) and (28,23) .. (nx32y19);
\draw[->,black] (nx32y11) -- (nx34y15);
\draw[->,black] (nx32y11) -- (nx30y15);
\draw[->,black] (nx32y19) -- (nx34y15);
\draw[->,black] (nx32y19) -- (nx30y15);
\draw[->,black] (nx24y19) -- (nx32y19);
\draw[->,black] (nx24y11) -- (nx32y11);
\draw[->,black] (nx26y15) -- (nx24y11);
\draw[->,black] (nx22y15) -- (nx24y11);
\draw[->,black] (nx26y15) -- (nx24y19);
\draw[->,black] (nx22y15) -- (nx24y19);
\draw[->,black] (nx8y19) -- (nx16y19);
\draw[->,black] (nx24y11) -- (nx16y11);
\draw[->,black] (nx24y19) -- (nx16y19);
\draw[->,black] (nx16y19) -- (nx18y15);
\draw[->,black] (nx16y19) -- (nx14y15);
\draw[->,black] (nx16y11) -- (nx14y15);
\draw[->,black] (nx16y11) -- (nx18y15);
\draw[->,black] (nx8y11) -- (nx16y11);
\draw[->,black] (nx10y15) -- (nx8y11);
\draw[->,black] (nx6y15) -- (nx8y11);
\draw[->,black] (nx10y15) -- (nx8y19);
\draw[->,black] (nx6y15) -- (nx8y19);
\draw (34,15) node[anchor = north west] {$8'$};
\draw (26,15) node[anchor = north west] {$8$};
\draw (18,15) node[anchor = north west] {$6'$};
\draw (10,15) node[anchor = north west] {$4$};
\draw (34,15) node[anchor = south west] {$4'$};
\draw (26,15) node[anchor = south west] {$5$};
\draw (18,15) node[anchor = south west] {$2'$};
\draw (10,15) node[anchor = south west] {$2$};
\draw (30,15) node[anchor = north east] {$7'$};
\draw (22,15) node[anchor = north east] {$7$};
\draw (14,15) node[anchor = north east] {$5'$};
\draw (6,15) node[anchor = north east] {$3$};
\draw (30,15) node[anchor = south east] {$3'$};
\draw (22,15) node[anchor = south east] {$6$};
\draw (14,15) node[anchor = south east] {$1'$};
\draw (6,15) node[anchor = south east] {$1$};
\end{tikzpicture}
\end{center}
\caption{Example of a planar acyclic instance with $r+c$ Eulerian, where there is a half-integral solution, but no integral solution.}
\label{fig:half-integral}
\end{figure}

When $G$ is an acyclic digraph and $|E(H)|$ is bounded, the problem is still NP-hard when $r+c$ is Eulerian (Vygen~\cite{vygen1995ncs}), or when $G$ is planar~\cite{naves08}. When $G+H$ is planar, the polynomiality of the integer multiflow problem follows from the theorem of Lucchesi-Younger~\cite{lucchesi1978mtd} on directed cuts packing. When $G$ is a planar acyclic digraph, $r+c$ is Eulerian but $|E(H)|$ is arbitrary, the problem is NP-hard (Marx~\cite{marx2004edp}). An example of an instance where there is a half-integral solution but no integral solution is given in Figure~\ref{fig:half-integral}. In contrast, if $r(E(H))$ is bounded, it is polynomially solvable (Fortune, Hopcroft and Wyllie~\cite{ForHopWyl80}), even without the planarity assumption. The following theorem states a pseudo-polynomiality result when $|E(H)|$ is bounded but not $r(E(H))$.

\begin{theorem}\label{theorem}
The arc-disjoint paths problem when $G$ is a planar acyclic digraph and $G+H$ is Eulerian, is decidable in time $O(f(R)^{k^2} k^3 n)$ where $f$ is a polynomial, $n = |V(G)|$, $k = |E(H)|$ and $R = \max_{h \in E(H)} r(h)$.
\end{theorem}

This result is partly inspired by Schrijver's work~\cite{schrijver1990hrm},~\cite{schrijver1993cdp} on vertex-disjoint paths problems in graphs embedded on surfaces. Vertex-disjoints paths on a surface can be represented as non-intersecting curves on the same surface. Schrijver's main idea is to guess the homotopies of those curves, and then solve the vertex-disjoint paths problem with these additional topologic constraints. As we are working with arc-disjoint paths, these can intersect in vertices, and the homotopic routing method does not work in our case. Nevertheless, we show that there is a canonical way for paths to relate to one another, and acyclicity restricts the number of guesses one has to provide in order to have a correct one among them.  

Theorem~\ref{theorem} can also be compared to a theorem of Ibaraki and Nagamochi~\cite{nagamochi1990multicommodity}, stating the polynomiality of the integer multiflow problem when $G$ is a planar acyclic digraph, $G+H$ is Eulerian, all the sources of $G$ and the tails of demand arcs are on the outer boundary of $G$. Their algorithm obeys a fully polynomial time bound and does not fix the number of demand arcs, at the price of a stronger assumption on $G$.


\section{Outline of the proof}

We now sketch the proof of Theorem~\ref{theorem}, before giving a more formal proof. Let $(G,H,r,c)$ be an Eulerian instance with $G$ planar acyclic. Our algorithm first guesses, for any pair of paths in the solution, what their behaviour is at their common vertices. Suppose that $P$ and $Q$ intersect at vertex $v$. Because we are dealing with planar graphs, we can describe the behaviour of $P$ in the following way: either $P$ crosses $Q$, or $P$ turns to the left, or $P$ turns to the right of $Q$. Considering every vertex and every pair of paths, this gives a behaviour pattern. We show that the behaviour pattern uniquely determines the solution to the arc-disjoint paths problem. In order to prove that, note that because $G+H$ is Eulerian and $G$ is acyclic, every arc is saturated by a solution (because the residual capacities define an Eulerian acyclic digraph, that is a graph with no arc). The following lemma is the key to the uniqueness:

\begin{lemma}\label{lemma:unique-routing}
Let $v$ be a vertex of $G$ with $n$ incoming arcs and $n$ outgoing arcs, and $\mathcal{P}$ a set of $n$ arc-disjoint paths containing $v$. Then the outgoing arc of each path is uniquely determined by the incoming arcs of each path going through $v$ and the relative behaviours of paths at $v$.
\end{lemma}

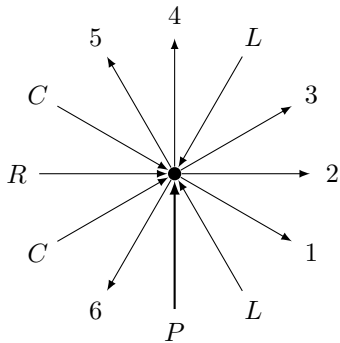
\begin{figure}[htb]
\begin{center}
\begin{tikzpicture}[x=0.6cm,y=0.6cm,>=latex]
\Vertex{black} (o) at (0,0) {};
\draw[->] (o) -- (0:3);\draw (0:3.5) node {$2$};
\draw[->] (o) -- (30:3);\draw (30:3.5) node {$3$};
\draw[<-] (o) -- (60:3);\draw (60:3.5) node {$L$};
\draw[->] (o) -- (90:3);\draw (90:3.5) node {$4$};
\draw[->] (o) -- (120:3);\draw (120:3.5) node {$5$};
\draw[<-] (o) -- (150:3);\draw (150:3.5) node {$C$};
\draw[<-] (o) -- (180:3);\draw (180:3.5) node {$R$};
\draw[<-] (o) -- (210:3);\draw (210:3.5) node {$C$};
\draw[->] (o) -- (240:3);\draw (240:3.5) node {$6$};
\draw[<-,thick] (o) -- (270:3);\draw (270:3.5) node {$P$};
\draw[<-] (o) -- (300:3);\draw (300:3.5) node {$L$};
\draw[->] (o) -- (330:3);\draw (330:3.5) node {$1$};
\end{tikzpicture}
\end{center}
\caption{An example of routing problem in a vertex.}
\label{fig:routing-a-vertex}
\end{figure}

Before proving the lemma, we explain what happens on the example of Figure~\ref{fig:routing-a-vertex}. Suppose the path $P$ (whose entering arc is indicated) must go to the left of the paths using arcs $L$, to the right of the one coming from $R$, and cross the paths from $C$. It cannot leave by arcs $1$, $2$ and $3$, without violating the condition on paths from $L$ arcs. It cannot leave by arc $6$ because of path $R$. If it leaves by arc $4$, then the two paths $L$ and the two paths $C$ must leave by arcs $1$, $2$ and $3$. Hence $P$ can only leave by arc $5$.

\begin{proof}
Let $P$ be a path entering $v$. Let $C$, $L$ and $R$ be respectively the sets of paths crossing $P$, going to the left of $P$ and to the right of $P$ at $v$. Let $a_0,a_1,\ldots,a_{2k-1}$ be the arcs incident to $v$, occurring in that order around $v$, with $a_0$ being the incoming arc of $P$.

Let $i$ be such that $P$ leaves $v$ by $a_i$. The paths in $L$ must have their arcs incident to $v$ in $I_1 + \{a_1,\ldots,a_{i-1}\}$ by definition of left. Similarly, the paths in $R$ use arcs in $I_2 = \{a_{i+1},\ldots,a_{2k-1}\}$. The paths in $C$ have one arc in $I_1$ and one in $I_2$. Thus $i = 2 |L| + |C| + 1$ is the unique possible candidate. 




By repeating the argument for each path, the solution is unique.

Note that to be feasible, for each path we must have that $a_i$ is a leaving arc that is between the last incoming arc of a path in $L$ and the first incoming arc of a path in $R$.
\end{proof}

To get the uniqueness for the whole graph, it is then sufficient to consider the vertices in an acyclic ordering, and route them consecutively (see the end of Section~\ref{sec:proof} for the detailed algorithm). It proves that we only need to know the relative behaviour of the paths. Guessing naively the behaviour pattern would lead to an exponential algorithm. The main part of the proof is then to show how we can drastically reduce the quantity of information needed to guess the behaviours of paths. We prove the following lemma:

\begin{lemma}\label{lemma:routing-schemes}
There is a function $F : \nat \times \nat \rightarrow \nat$ such that for every network $(G,H,r,c)$ with $G$ planar acyclic and $G+H$ Eulerian, it is sufficient to consider only $O(F(R,k))$ behaviour patterns, where $R = \max_{h \in E(H)} r(h)$,  $k = |E(H)|$.
\end{lemma} 

Here, the important fact is that the number of routing schemes does not depend on $G$. We will also show that these patterns can be enumerated. The rest of the paper is dedicated to the proof of this lemma.

\section{Notations and terminology}

Recall that $G$ is a planar acyclic supply digraph, $H$ is a demand graph, and $G+H$ is Eulerian. We assume that an embedding of $G$ on the sphere is given. The capacity of every arc is $1$. This can be achieved by replicating each arc in as many copies as its capacity. Without loss of generality, the tail of each demand arc is a sink and its head is a source of $G$; for a demand arc $vu$, add two new terminals $s_{uv}$ and $t_{uv}$, $r(vu)$ supply arcs $s_{uv}u$ and $vt_{uv}$, and a demand arc $t_{uv}s_{uv}$ with request $r(vu)$. So $H$ can be supposed to be a matching.

A \emph{path} $P$ is a sequence of distinct arcs such that the head of an arc is the tail of the following arc. The \emph{origin} of the path is the tail of the first arc, and the \emph{destination} is the head of the last arc. Since we work with acyclic digraphs, our paths are always simple. We denote $E(P)$ the arcs of $P$, and $V(P)$ the vertices incident to $E(P)$. The origin and the destination of a path are its \emph{endpoints}, all the other vertices of $V(P)$ are said to be \emph{internal} If $P$ is a path with destination $u$, and $Q$ is a path arc-disjoint form $P$ with origin $u$, we denote $PQ$ the concatenation of the two paths, that is the path $R$ with the same origin as $P$, the same destination as $Q$, and $E(R) = E(P) \cup E(Q)$. $P^{-1}$ is the path obtained by reversing each arc of $P$ (this notation is abusive, as $P^{-1}$ is not a path in $G$ but in a graph obtained from $G$ by reversing some arcs, but we will use this notation only for defining undirected cycle of $G$). If $u$ and $v$ are in $V(P)$, we denote $P_{uv}$ the subpath of $P$ with origin $u$ and destination $v$, $P_{\bot{}u}$ the subpath with the same origin as $P$ and destination $u$, and $P_{u\top}$ the subpath with origin $u$ and the same destination as $P$. A path with origin $s$ and destination $t$ is called an $(s,t)$-path.

We introduce basic tools to work on the topology of paths in the plane. Let $e_1,\ldots,e_n$ be any family indexed by the interval $\llbracket 1; n\rrbracket$. Elements $(e_i, e_j)$ are said to be \emph{consecutive} if $j = i+1$, or $i=n$ and $j=1$, thus defining a \emph{cyclic order}. Let $a$, $b$ and $c$ be three elements in this family, $b$ is \emph{between} $a$ and $c$ if $b$ appears in the sequence of consecutive elements $a = u_1, u_2,\ldots ,u_k=c$, and \emph{strictly between} if $b \notin \{a,c\}$. If $b$ is strictly between $a$ and $c$, $b$ is not between $c$ and $a$. An \emph{interval} of a cyclic order is either the empty set, or the set of elements between a given pair of elements which are then called the \emph{extremities} of the interval. Note that any pair defines two distinct intervals, depending on their order. The complement of an interval is an interval. We say that $b$ is \emph{before} $c$ in the interval of extremities $(a,d)$ if $b$ and $c$ are between $a$ and $d$ and $b$ is between $a$ and $c$.

In a planar digraph, the order of appearance of the arcs incident to a vertex $v$ induces a cyclic order; $y$ is between  $x$ and $z$ if $x$, $y$ and $z$ appears in that order around $v$ in the positive (anticlockwise) orientation. Two consecutive arcs are adjacent to a common face. This order induces two cycle orders on the incoming and on the outgoing arcs of $v$. Hence we have three orders, for $\delta(v)$, $\delta^+(v)$ and $\delta^-(v)$ respectively. Any set of arc-disjoint paths sharing the same endpoints will be considered with the cyclic order defined by their first arcs leaving their common source. All the indices in the following will be chosen with respect to these cyclic orders, and will be considered modulo the cardinality of the family.

\begin{figure}
\begin{center}
\begin{tikzpicture}[x=0.5cm,y=0.5cm,>=latex]
\vertex{black} (nx23y18) at (23,18) {};
\vertex{black} (nx23y14) at (23,14) {};
\vertex{black} (nx19y18) at (19,18) {};
\vertex{black} (nx19y14) at (19,14) {};
\vertex{black} (nx16y14) at (16,14) {};
\vertex{black} (nx16y18) at (16,18) {};
\vertex{black} (nx12y18) at (12,18) {};
\vertex{black} (nx12y14) at (12,14) {};
\vertex{black} (nx9y14) at (9,14) {};
\vertex{black} (nx5y18) at (5,18) {};
\vertex{black} (nx5y14) at (5,14) {};
\vertex{black} (nx9y18) at (9,18) {};

\Vertex{black} (nx7y16) at (7,16) {};
\Vertex{black} (nx21y16) at (21,16) {};
\Vertex{black} (nx14y16) at (14,16) {};

\draw (7,16) node[anchor=south] {$v$};
\draw (14,16) node[anchor=south] {$v$};
\draw (21,16) node[anchor=south] {$v$};

\draw[->,black] (nx14y16) -- (nx12y14);
\draw[->,black] (nx16y14) -- (nx14y16);
\draw[dashed] (nx21y16) -- (nx23y18);
\draw[dashed] (nx19y18) -- (nx21y16);
\draw[dashed] (nx14y16) -- (nx16y18);
\draw[dashed] (nx12y18) -- (nx14y16);
\draw[->,black] (nx21y16) -- (nx23y14);
\draw[->,black] (nx19y14) -- (nx21y16);
\draw[->,dashed] (nx9y14) -- (nx7y16);
\draw[->,dashed] (nx7y16) -- (nx5y18);
\draw[->,black] (nx5y14) -- (nx7y16);
\draw[->,black] (nx7y16) -- (nx9y18);
\draw (21,12) node {To the right};
\draw (14,12) node {To the left};
\draw (7,12) node {Crossing};
\draw (9,18) node[anchor = south west] {$P^+_v$};
\draw (23,14) node[anchor = north west] {$P^+_v$};
\draw (16,14) node[anchor = north west] {$P^-_v$};
\draw (19,14) node[anchor = north east] {$P^-_v$};
\draw (12,14) node[anchor = north east] {$P^+_v$};
\draw (5,14) node[anchor = north east] {$P^-_v$};
\draw (22,18) node {$Q$};
\draw (15,18) node {$Q$};
\draw (5,18) node[anchor = south east] {$Q^+_v$};
\draw (9,14) node[anchor = north west] {$Q^+_v$};
\end{tikzpicture}
\caption{The behaviour of $P$ (plain) at $v$ relative to $Q$ (dashed).}
\label{fig:crossing}
\end{center}

\end{figure}
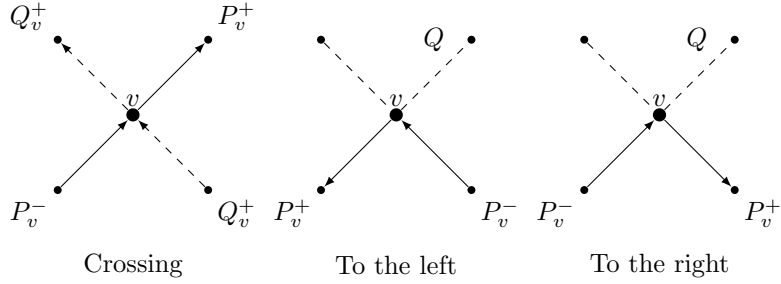

For any path $P$ and any vertex $v \in V(P)$, we denote $P_v^-$ and $P_v^+$ the arcs of $P$ entering and leaving $v$, when they exist. Let $P$ and $Q$ be two distinct paths, and $v$ be an inner vertex of both $P$ and $Q$. We say that (see Figure~\ref{fig:crossing}):
\begin{itemize}
\item $P$ and $Q$ \emph{cross at} $v$ if $P_v^-$ and $P_v^+$ are not in the same interval with extremities $Q_v^-$ and $Q_v^+$ (in the cyclic order induced by $\delta(v)$),
\item $P$ \emph{goes to the right} of $Q$ at $v$ if $P_v^-$ and $P_v^+$ are  in the same interval with extremities $Q_v^-$ and $Q_v^+$, and $P_v^-$ is before $P_v^+$ in this interval.
\item $P$ \emph{goes to the left} of $Q$ at $v$ if  $P_v^-$ and $P_v^+$ are  in the same interval with extremities $Q_v^-$ and $Q_v^+$, and $P_v^+$ is before $P_v^-$ in this interval.
\end{itemize}

$P$ and $Q$ \emph{cross} if they cross at $v$ for some vertex $v$, otherwise they are \emph{uncrossed}. From the definition, the following statements are equivalent:
\begin{itemize}
\item[$(i)$] $P$ goes to the left of $Q$ at $v$,
\item[$(ii)$] $P$ goes to the left of $Q^{-1}$ at $v$,
\item[$(iii)$] $P^{-1}$ goes to the right of $Q$ at $v$.
\end{itemize} 
The knowledge of whether $P$ crosses, goes to the left or to the right of $Q$ at $v$ will be called the \emph{behaviour} of $P$ relative to $Q$ at $v$. One of the main ideas of our proof is that the solutions are determined by the relative behaviours of each pair of paths at each node.

An \emph{acyclic} (or \emph{topological}) order for $G$ is a linear order of the vertices such that for all arcs $uv \in E(G)$, $v$ is greater than $u$. A digraph admits an acyclic order if and only if it is acyclic. We assume that we are given an acyclic order $<$ for $G$. We call the \emph{first common vertex} of $P$ and $Q$ the smallest vertex of $V(P) \cap V(Q)$.

\section{Proof of the theorem}\label{sec:proof}

\begin{lemma}\label{lemma:uncrossing}
Let $G$ be a planar acyclic digraph and $\mathcal{P}$ be a set of arc-disjoint paths in $G$. Then there is a set $\mathcal{P}'$ of arc-disjoint paths satisfying the same demands as $\mathcal{P}$, such that:
\begin{itemize}
\item[-] two paths of $\mathcal{P}'$ sharing a common endpoint do not cross,
\item[-] two crossed paths of $\mathcal{P}'$ cross at a unique vertex, their first common vertex.
\end{itemize}
\end{lemma}

\begin{proof}
Let $P$ and $Q$ be two paths of $\mathcal{P}$, and $u < v$ be two vertices in $V(P) \cap V(Q)$. Suppose that $P$ and $Q$ cross at $v$, or cross at $u$ and $v$ is their common destination (if there are no $P$ and $Q$ with these properties, we choose $\mathcal{P}' = \mathcal{P}$).

We define $P'=P_{\bot{}u}Q_{uv}P_{v\top}$ and $Q'=Q_{\bot{}u}P_{uv}Q_{v\top}$. Let $\mathcal{P}' = (\mathcal{P} \setminus \{P,Q\}) \cup \{P',Q'\}$. Note that $\mathcal{P'}$ satisfies the same demands as $\mathcal{P}$. For any vertex $w$ different from $u$ and $v$, the number of pairs of paths that cross at $w$ does not change. if $P$ and $Q$ cross at $v$ and not at $u$, then the number of crossings at $v$ is decreased by at least $1$. Otherwise, the number of crossing at $u$ is decreased by at least $1$, while the number of crossing at $v$ is not increased. Consider for a set of arc-disjoint paths $\mathcal{Q}$ the vector $\cross(\mathcal{Q}) \in \mathbb{N}^{|V|}$, whose $i$th coordinate is the number of crossings in $\mathcal{Q}$ at the $i^{\textrm{th}}$ greatest vertex. Then this vector is lexicographically decreased by the transformation, $\cross(\mathcal{P'}) < \cross(\mathcal{P})$. By induction on this vector, the lemma is proved. 
\end{proof}

We remark that Ibaraki and Nagamochi's theorem~\cite{nagamochi1990multicommodity} (discussed at the end of the introduction) is a consequence of this lemma. Let's state their theorem.

\begin{theorem}[Ibaraki and Nagamochi, 1990]
There is a polynomial-time algorithm to solve the arc-disjoint paths problem when $G$ is a planar directed acyclic graph, $r+c$ is Eulerian, the tails of the demand arcs and the sources of $G$ are on the boundary of the outer face of $G$.
\end{theorem}

\begin{proof}
Take a source $s$ on the outer boundary of $G$. Because $G+H$ is Eulerian, we
have $|\delta^+(s)|$ demand arcs with destination $s$. The corresponding
paths, if they exists, can be chosen uncrossed by
Lemma~\ref{lemma:uncrossing}. Now we can order the demand arcs depending on
the order of their tails on the outer boundary of $G$, this determines the
order used by the corresponding paths to leave $s$. Then, we can reduce the
original problem to a new problem on $G \setminus \{v\}$, satisfying the same
assumptions (new sources are also on the outer boundary). Deriving a
polynomial-time is then straightforward.
\end{proof}

A set of arc-disjoint paths is said to be \emph{uncrossed} if it satisfies the two conditions of Lemma~\ref{lemma:uncrossing}. We only consider uncrossed sets.

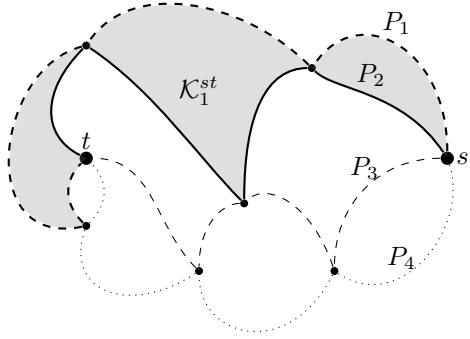
\begin{figure}
\begin{center}
\begin{tikzpicture}[x=0.3cm,y=0.3cm,>=latex,rotate=90]
\vertex{black} (nx13y16) at (13,16) {};\point{} (nx14y15) at (14,15) {};
\point{} (nx18y20) at (18,20) {};\point{} (nx15y18) at (15,18) {};
\point{} (nx16y25) at (16,25) {};\point{} (nx18y25) at (18,25) {};
\vertex{black} (nx20y23) at (20,23) {};\point{} (nx11y28) at (11,28) {};
\point{} (nx14y24) at (14,24) {};\point{} (nx13y24) at (13,24) {};
\point{} (nx19y27) at (19,27) {};\point{} (nx23y21) at (23,21) {};
\point{} (nx22y16) at (22,16) {};\point{} (nx20y7) at (20,7) {};
\point{} (nx22y11) at (22,11) {};\vertex{black} (nx19y13) at (19,13) {};
\point{} (nx17y16) at (17,16) {};\point{} (nx19y15) at (19,15) {};
\point{} (nx18y12) at (18,12) {};\point{} (nx18y9) at (18,9) {};
\point{} (nx12y20) at (12,20) {};\point{} (nx15y20) at (15,20) {};
\point{} (nx12y18) at (12,18) {};\point{} (nx13y17) at (13,17) {};
\point{} (nx13y13) at (13,13) {};\point{} (nx13y12) at (13,12) {};
\point{} (nx15y10) at (15,10) {};\vertex{black} (nx12y23) at (12,23) {};
\vertex{black} (nx10y18) at (10,18) {};\vertex{black} (nx10y12) at (10,12) {};
\point{} (nx14y22) at (14,22) {};\point{} (nx13y22) at (13,22) {};
\point{} (nx9y24) at (9,24) {};\point{} (nx8y21) at (8,21) {};
\point{} (nx7y18) at (7,18) {};\point{} (nx6y14) at (6,14) {};
\point{} (nx8y9) at (8,9) {};\point{} (nx12y6) at (12,6) {};

\Vertex{black} (nx15y23) at (15,23) {};\Vertex{black} (nx15y7) at (15,7) {};
\draw (15,7) node[anchor=west] {$s$};
\draw (15,23) node[anchor=south] {$t$};

\fill[fill=gray,nearly transparent]
   (15,7) .. controls (18,9) and (18,12) .. (19,13)
.. controls (19,15) and (17,16) .. (13,16)
.. controls (15,18) and (18,20) .. (20,23)
.. controls (18,25) and (16,25) .. (15,23)
.. controls (14,24) and (13,24) .. (12,23)
.. controls (11,28) and (19,27) .. (20,23)
.. controls (23,21) and (22,16) .. (19,13)
.. controls (22,11) and (20,7) .. (15,7);
\draw (18,18) node {$\mathcal{K}_1^{st}$};

\draw[black,thick,dashed] (nx12y23) .. controls (nx13y24) and (nx14y24) .. (nx15y23);
\draw[black,thick,dashed] (nx20y23) .. controls (nx19y27) and (nx11y28) .. (nx12y23);
\draw[black,thick,dashed] (nx19y13) .. controls (nx22y16) and (nx23y21) .. (nx20y23);
\draw[black,thick,dashed] (nx15y7) .. controls (nx20y7) and (nx22y11) .. (nx19y13) 
  node[midway,above] {$P_1$};

\draw[black,thick] (nx20y23) .. controls (nx18y25) and (nx16y25) .. (nx15y23);
\draw[black,thick] (nx13y16) .. controls (nx15y18) and (nx18y20) .. (nx20y23);
\draw[black,thick] (nx19y13) .. controls (nx19y15) and (nx17y16) .. (nx13y16);
\draw[black,thick] (nx15y7) .. controls (nx18y9) and (nx18y12) .. (nx19y13) 
  node[midway,above] {$P_2$};

\draw[black,dashed] (nx10y18) .. controls (nx12y20) and (nx15y20) .. (nx15y23);
\draw[black,dashed] (nx13y16) .. controls (nx13y17) and (nx12y18) .. (nx10y18);
\draw[black,dashed] (nx10y12) .. controls (nx13y13) and (nx14y15) .. (nx13y16);
\draw[black,dashed] (nx15y7) .. controls (nx15y10) and (nx13y12) .. (nx10y12) 
  node[midway,above] {$P_3$};

\draw[black,dotted] (nx12y23) .. controls (nx13y22) and (nx14y22) .. (nx15y23);
\draw[black,dotted] (nx10y18) .. controls (nx8y21) and (nx9y24) .. (nx12y23);
\draw[black,dotted] (nx10y12) .. controls (nx6y14) and (nx7y18) .. (nx10y18);
\draw[black,dotted] (nx15y7) .. controls (nx12y6) and (nx8y9) .. (nx10y12) 
  node[midway,left] {$P_4$};

\end{tikzpicture}
\end{center}
\caption{One of the four regions defined by the four paths of a commodity}
\label{fig:regions}
\end{figure}

Recall that $G$ is embedded on the sphere, and that two disjoint-paths with the same extremities are consecutive if their first arcs are consecutive. Let $P$ and $Q$ be two uncrossed arc-disjoint $(s,t)$-paths. Because $P$ and $Q$ are arc-disjoint, $PQ^{-1}$ defines a closed curve on the sphere, which is simple within an arbitrarily small continuous deformation (because of the potential common inner vertices). By Jordan's theorem, this simple curve separates the sphere in two connected components. We call \emph{inside} (or \emph{interior}) the component for which $PQ^{-1}$ runs positively along its boundary, and \emph{outside} the other component. If $P$ and $Q$ are consecutive, any other $(s,t)$-path lies on the outside of $PQ^{-1}$. Thus, given a solution, for any demand arc $h = ts$, we can partition the sphere in $r(h)$ disjoint parts, called \emph{regions}, which are the interiors of two consecutive $(s,t)$-paths. We denote $\mathcal{K}_i^h$ the region defined by the $P_iP^{-1}_{i+1}$. The regions associated with a demand are considered in the cyclic order induced by their indices. Figure~\ref{fig:regions} shows the region $\mathcal{K}_1^{st}$ associated with the four paths of a commodity $st$. 

\begin{lemma}\label{lemma:crossing-intervals}
Let $t_1s_1, t_2s_2$ be two distinct demand arcs, with respective requests $r_1$ and $r_2$, and $\mathcal{P}$ be an uncrossed solution. Then there is an interval $\mathcal{I}_1$ of $(s_1,t_1)$-paths and an interval $\mathcal{I}_2$ of $(s_2,t_2)$-paths in $\mathcal{P}$, such that for any $(s_1,t_1)$-path $P$ and any $(s_2,t_2)$-path $Q$ in $\mathcal{P}$, $P$ and $Q$ are crossed if and only if either $P \in \mathcal{I}_1$ and $Q \in \mathcal{I}_2$, or $P \notin \mathcal{I}_1$ and $Q \notin \mathcal{I}_2$.
\end{lemma}

\begin{figure}
\begin{center}
\begin{tikzpicture}[x=0.3cm,y=0.3cm,>=latex,rotate=90]
\point{} (nx13y16) at (13,16) {};\point{} (nx14y15) at (14,15) {};
\point{} (nx18y20) at (18,20) {};\point{} (nx15y18) at (15,18) {};
\point{} (nx16y25) at (16,25) {};\point{} (nx18y25) at (18,25) {};
\point{} (nx20y23) at (20,23) {};\point{} (nx11y28) at (11,28) {};
\point{} (nx14y24) at (14,24) {};\point{} (nx13y24) at (13,24) {};
\point{} (nx19y27) at (19,27) {};\point{} (nx23y21) at (23,21) {};
\point{} (nx22y16) at (22,16) {};\point{} (nx20y7) at (20,7) {};
\point{} (nx22y11) at (22,11) {};\point{} (nx19y13) at (19,13) {};
\point{} (nx17y16) at (17,16) {};\point{} (nx19y15) at (19,15) {};
\point{} (nx18y12) at (18,12) {};\point{} (nx18y9) at (18,9) {};
\point{} (nx12y20) at (12,20) {};\point{} (nx15y20) at (15,20) {};
\point{} (nx12y18) at (12,18) {};\point{} (nx13y17) at (13,17) {};
\point{} (nx13y13) at (13,13) {};\point{} (nx13y12) at (13,12) {};
\point{} (nx15y10) at (15,10) {};\point{} (nx12y23) at (12,23) {};
\point{} (nx10y18) at (10,18) {};\point{} (nx10y12) at (10,12) {};
\point{} (nx14y22) at (14,22) {};\point{} (nx13y22) at (13,22) {};
\point{} (nx9y24) at (9,24) {};\point{} (nx8y21) at (8,21) {};
\point{} (nx7y18) at (7,18) {};\point{} (nx6y14) at (6,14) {};
\point{} (nx8y9) at (8,9) {};\point{} (nx12y6) at (12,6) {};

\Vertex{black} (nx15y23) at (15,23) {};\Vertex{black} (nx15y7) at (15,7) {};
\draw (15,23) node[anchor=south] {$s_1$};
\draw (15,7) node[anchor=west] {$t_1$};

\fill[gray,nearly transparent]
   (15,7) .. controls (18,9) and (18,12) .. (19,13)
.. controls (19,15) and (17,16) .. (13,16)
.. controls (15,18) and (18,20) .. (20,23)
.. controls (18,25) and (16,25) .. (15,23)
.. controls (14,24) and (13,24) .. (12,23)
.. controls (11,28) and (19,27) .. (20,23)
.. controls (23,21) and (22,16) .. (19,13)
.. controls (22,11) and (20,7) .. (15,7);
\fill[gray,nearly transparent] 
  (15,7) .. controls (12,6) and (8,9) .. (10,12)
.. controls (6,14) and (7,18) .. (10,18)
.. controls (8,21) and (9,24) .. (12,23)
.. controls (13,22) and (14,22) .. (15,23)
.. controls (15,20) and  (12,20) .. (10,18)
.. controls (12,18) and (13,17) .. (13,16)
.. controls (14,15) and (13,13) .. (10,12)
.. controls (13,12) and (15,10) .. (15,7);

\draw[dashed] (nx12y23) .. controls (nx13y24) and (nx14y24) .. (nx15y23);
\draw[dashed] (nx20y23) .. controls (nx19y27) and (nx11y28) .. (nx12y23);
\draw[dashed] (nx19y13) .. controls (nx22y16) and (nx23y21) .. (nx20y23);
\draw[dashed] (nx15y7) .. controls (nx20y7) and (nx22y11) .. (nx19y13);

\draw[black] (nx20y23) .. controls (nx18y25) and (nx16y25) .. (nx15y23);
\draw[black] (nx13y16) .. controls (nx15y18) and (nx18y20) .. (nx20y23);
\draw[black] (nx19y13) .. controls (nx19y15) and (nx17y16) .. (nx13y16);
\draw[black] (nx15y7) .. controls (nx18y9) and (nx18y12) .. (nx19y13);

\draw[black] (nx10y18) .. controls (nx12y20) and (nx15y20) .. (nx15y23);
\draw[black] (nx13y16) .. controls (nx13y17) and (nx12y18) .. (nx10y18);
\draw[black] (nx10y12) .. controls (nx13y13) and (nx14y15) .. (nx13y16);
\draw[black] (nx15y7) .. controls (nx15y10) and (nx13y12) .. (nx10y12);

\draw[black,dashed] (nx12y23) .. controls (nx13y22) and (nx14y22) .. (nx15y23);
\draw[black,dashed] (nx10y18) .. controls (nx8y21) and (nx9y24) .. (nx12y23);
\draw[black,dashed] (nx10y12) .. controls (nx6y14) and (nx7y18) .. (nx10y18);
\draw[black,dashed] (nx15y7) .. controls (nx12y6) and (nx8y9) .. (nx10y12);

\Vertex{black} (nx21y19) at (21,19) {};\draw (21,19) node[anchor=north west] {$s_2$};
\Vertex{black} (nx9y15) at (9,15) {};\draw (9,15) node[anchor=north west] {$t_2$};

\draw[black] (nx19y13) .. controls (21,16) and (21,18) .. (nx21y19);
\draw[black] (nx9y15) .. controls (11,13) and (16,12) .. (nx19y13);

\draw[black] (nx13y16) .. controls (14,19) and (16,20) .. (nx21y19);
\draw[black] (nx9y15) .. controls (10,16) and (12,15) .. (nx13y16);

\draw[black] (nx10y18) .. controls (10,21) and (17,22) .. (nx21y19);
\draw[black] (nx9y15) .. controls (9,16) and (9,17) .. (nx10y18);

\draw[black,dashed] (22,28) .. controls (23,26) and (23,22) .. (nx21y19);
\draw[black,dashed] (13,32) .. controls (17,32) and (21,30) .. (22,28);
\draw[black,dashed] (4,23) .. controls (4,27) and (9,32) .. (13,32);
\draw[black,dashed] (nx9y15) .. controls (7,15) and (4,19) .. (4,23);

\draw[black,dashed] (nx20y23) .. controls (21,21) and (21,20) .. (nx21y19);
\draw[black,dashed] (10,29) .. controls (13,31) and (19,29) .. (nx20y23);
\draw[black,dashed] (nx9y15) .. controls (6,21) and (7,27) .. (10,29);

\draw[black,dashed] (16,4) .. controls (20,4) and (25,12) .. (nx21y19);
\draw[black,dashed] (nx10y12) .. controls (6,9) and (12,4) .. (16,4);
\draw[black,dashed] (nx9y15) .. controls (10,14) and (10,13) .. (nx10y12);

\end{tikzpicture}
\end{center}

\caption{The dashed $(s_1,t_1)$-paths crosses the dashed $(s_2,t_2)$-paths and are consecutive around their sources.}
\label{fig:intervals}
\end{figure}
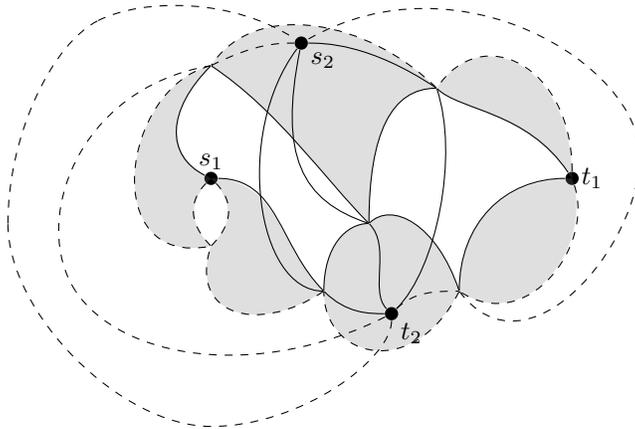

Figure~\ref{fig:intervals} illustrates this lemma. The two intervals are given by the dashed paths.Two paths from two distinct commodities cross each other if and only if they are both dashed, or both plain.

\begin{proof}
Let $P_1,\ldots,P_{r_1}$ be the paths satisfying the demand $t_1s_1$. $s_2$ and $t_2$ are contained in $\mathcal{K}_i^{t_1s_1}$ and  $\mathcal{K}_j^{t_1s_1}$ respectively. Because $s_1$ and $t_1$ are a source and a sink of $G$, any $(s_2,t_2)$-path must intersect consecutive regions of $t_1s_1$. As two consecutive regions are separated by a $(s_1,t_1)$-path, a path going from region $\mathcal{K}_i^{t_1s_1}$ to $\mathcal{K}_{i+1}^{t_1s_1}$ crosses $P_i$. Because the solution is uncrossed, an $(s_2,t_2)$-path must go through a monotonic sequence of regions, either $\mathcal{K}_i^{t_1s_1},\mathcal{K}_{i+1}^{t_1s_1},\ldots,\mathcal{K}_j^{t_1s_1}$ or $\mathcal{K}_i^{t_1s_1},\mathcal{K}_{i-1}^{t_1s_1},\ldots,\mathcal{K}_j^{t_1s_1}$.This defines $\mathcal{I}_1 = P_{i+1} \cup P_{i+2} \cup \ldots \cup P_j$, and $\mathcal{I}_2$ the sets of paths crossing the paths in $\mathcal{I}_1$. By symmetry, $\mathcal{I}_2$ is also an interval.
\end{proof}

\begin{lemma}\label{lemma:second-intersection}
Let $P$ and $Q$ be two paths of an uncrossed solution, $u < v$ be two vertices in $V(P) \cap V(Q)$. $P$ goes to the left of $Q$ at $v$ if and only if the destination of $P$ is inside $P_{uv}Q^{-1}_{uv}$.
\end{lemma}

\begin{figure}
\begin{center}
\begin{tikzpicture}[x=0.3cm,y=0.3cm,>=latex,rotate=315]
\point{black} (nx18y23) at (18,23) {};
\point{black} (nx15y20) at (15,20) {};
\point{black} (nx10y17) at (10,17) {};
\point{black} (nx7y15) at (7,15) {};
\point{black} (nx16y12) at (16,12) {};
\point{black} (nx12y9) at (12,9) {};
\point{black} (nx24y18) at (24,18) {};
\point{black} (nx20y15) at (20,15) {};

\Vertex{black} (nx20y19) at (20,19) {};\draw (20,19) node[anchor=east] {$t_2$};
\Vertex{black} (nx6y7) at (6,7) {};\draw (6,7) node[anchor=east] {$s_2$};

\Vertex{black} (nx23y22) at (23,22) {};\draw (23,22) node[anchor=north west] {$w$};
\Vertex{black} (nx15y15) at (15,15) {};\draw (15,15) node[anchor=east] {$v$};
\Vertex{black} (nx8y11) at (8,11) {};\draw (8,11) node[anchor=west] {$u$};

\Vertex{black} (nx26y25) at (26,25) {};\draw (26,25) node[anchor=west] {$t_1$};
\Vertex{black} (nx3y12) at (3,12) {};\draw (3,12) node[anchor=east] {$s_1$};

\draw[->,black] (nx23y22) -- (nx20y19);
\draw[->,black] (nx15y15) .. controls (nx15y20) and (nx18y23) .. (nx23y22);
\draw[->,black] (nx8y11) .. controls (nx7y15) and (nx10y17) .. (nx15y15);
\draw[->,black] (nx6y7) -- (nx8y11);

\fill[black,semitransparent] 
  (8,11) .. controls (7,15) and (10,17) ..
  (15,15) .. controls (16,12) and (12,9) .. (8,11);
\fill[gray,nearly transparent]
  (15,15) .. controls (15,20) and (18,23) ..
  (23,22) .. controls (24,18) and (20,15) .. (15,15);

\draw[->,dashed] (nx23y22) -- (nx26y25);
\draw[->,dashed] (nx15y15) .. controls (nx20y15) and (nx24y18) .. (nx23y22);
\draw[->,dashed] (nx8y11) .. controls (nx12y9) and (nx16y12) .. (nx15y15);
\draw[->,dashed] (nx3y12) -- (nx8y11);
\end{tikzpicture}
\end{center}
\caption{The $(s_2,t_2)$-path $P_2$ goes to the left of the dashed path $P_1$ at $v$, because $t_2$ is inside the cycle $(P_2)_{uv}(P_1)_{uv}^{-1}$ (the outside is in dark grey), but goes to the right of $P_1$ at $w$ because $t_2$ is outside of $(P_2)_{vw}(P_1)_{vw}^{-1}$ (in light cray).}
\label{fig:second}
\end{figure}
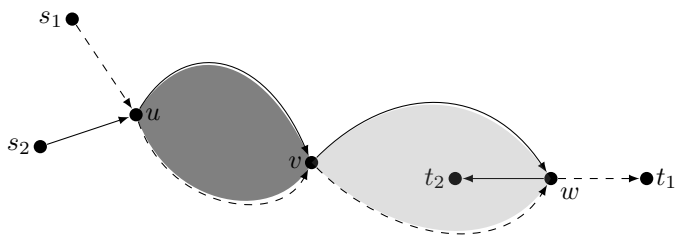

\begin{proof}
Note that $P$ and $Q$ do not cross at $v$. $P_{v\top}$ cannot intersect $P_{uv}Q_{uv}^{-1}$ by acyclicity, it is contained in the interior or the outside of $P_{uv}Q_{uv}^{-1}$. Thus the position of the destination of $P$ determines the behaviour of $P$ relatively to $Q$ at $v$ (see Figure~\ref{fig:second}).
\end{proof}

Thus, the relative behaviours of two paths at a vertex that is not their first common vertex depends only on the position of their terminals and the subpaths from their origin to the given vertex.

\begin{lemma}\label{lemma:two-cycles}
Let $C$ and $D$ be two cycles of a planar graph with disjoint interiors, and $v \in V(C)$ be a vertex that appears exactly once in $D$. Then $D$ goes to the left of $C$ at $v$.
\end{lemma}

\begin{proof}
If $D$ goes to the right of $C$ at $v$, $C$ meets the interior of $D$, contradiction.
\end{proof}

\begin{lemma}\label{lemma:non-crossing-intervals}
Let $P_1,\ldots,P_n$ be the $(s,t)$-paths of a solution, $Q$ and $R$ two paths for another request, and consider the cycle $C = QR^{-1}$. The set of $(s,t)$-paths that go to the left (resp. right) of $C$ on their first common vertex is an interval.
\end{lemma}

\begin{figure}
\begin{tabular}{cc}
\begin{minipage}{0.45\textwidth}
\begin{center}
\begin{tikzpicture}[x=0.6cm,y=0.6cm,>=latex]
\fill[gray,nearly transparent] (-2,-3) .. controls (0,-2) .. (260:1);
\fill[gray,nearly transparent] (-2,-3) .. controls (-2.5,-1.5) .. (-1,0);
\fill[gray,nearly transparent,even odd rule]
  (260:1) -- (-1,0) -- (-2,-3)
  (260:1) .. controls (220:1.5) .. (-1,0) -- cycle
;

\fill[black,semitransparent] (0,0) circle (1);

\draw[->] (1,0) arc (0:120:1);
\draw[->] (120:1) arc (120:240:1);
\draw[->] (240:1) arc (240:360:1);
\vertex{black} (v) at (-1,0) {};
\vertex{black} (u) at (260:1) {};
\Vertex{black} (s) at (-2,-3) {};
\Vertex{black} (t) at (-3,3) {};

\draw (0,0) node {$C$};
\draw (-1,-1.7) node {$D$};
\draw (260:1) node[anchor=north west] {$u$};
\draw (-1,0) node[anchor=west] {$v$};
\draw (s) node[anchor=north] {$s$};
\draw (t) node[anchor=east] {$t$};
\draw (-2.5,-1) node {$P_j$};
\draw (0,-2.5) node {$P_i$};

\draw[->,black] (s) .. controls (0,-2) .. (u);
\draw[->,black] (u) .. controls (220:1.5) .. (v);
\draw[->,black] (v) .. controls (-1,2) .. (t);

\draw[->,dashed,black] (s) .. controls (-2.5,-1.5) .. (v);
\draw[->,dashed,black] (v) .. controls (-3,1) .. (t); 
\end{tikzpicture}
\end{center}
\end{minipage}
&
\begin{minipage}{0.45\textwidth}
\begin{center}
\begin{tikzpicture}[x=0.6cm,y=0.6cm,>=latex]
\fill[gray,nearly transparent] (-2,-3) .. controls (-0.7,-2) .. (-1,0);
\fill[gray,nearly transparent] (-2,-3) .. controls (-2.5,-1.5) .. (-1,0);

\fill[black,semitransparent] (0,0) circle (1);

\fill[gray,nearly transparent]
  (-1,0) .. controls (-1,2) .. (-3,3) .. controls (-3,1) .. (-1,0); 

\draw[->] (1,0) arc (0:120:1);
\draw[->] (120:1) arc (120:240:1);
\draw[->] (240:1) arc (240:360:1);
\vertex{black} (v) at (-1,0) {};
\vertex{black} (u) at (260:1) {};
\Vertex{black} (s) at (-2,-3) {};
\Vertex{black} (t) at (-3,3) {};

\draw (0,0) node {$C$};
\draw (-1.4,-1.7) node {$D$};
\draw (-2,1.5) node {$D'$};
\draw (-1,0) node[anchor=west] {$v$};
\draw (s) node[anchor=north] {$s$};
\draw (t) node[anchor=east] {$t$};
\draw (-2.5,-1) node {$P_j$};
\draw (-0.7,-2.5) node {$P_i$};

\draw[->,black] (s) .. controls (-0.7,-2) .. (v);
\draw[->,black] (v) .. controls (-1,2) .. (t);

\draw[->,dashed,black] (s) .. controls (-2.5,-1.5) .. (v);
\draw[->,dashed,black] (v) .. controls (-3,1) .. (t); 
\end{tikzpicture}
\end{center}
\end{minipage}
\\
$(a)$ & $(b)$
\end{tabular}
\caption{Illustrations for Lemma~\ref{lemma:non-crossing-intervals}}
\label{fig:left-paths}
\end{figure}
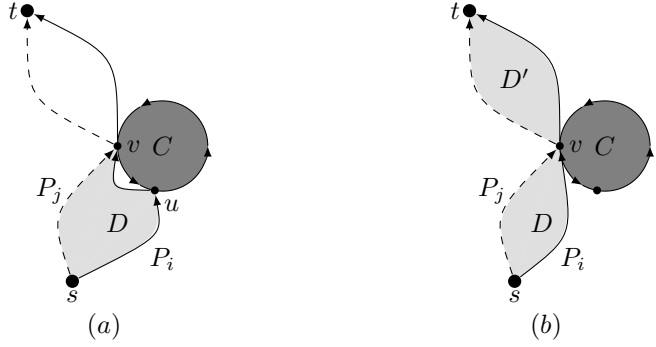

\begin{proof}
Without loss of generality, we index the $(s,t)$-paths such that the interior of $C$ is contained in $\mathcal{K}_k^{ts} \cup \ldots \cup \mathcal{K}_n^{ts}$ with $k$ maximal (thus $P_l$ goes inside $C$ if and only if $k < l \leq n$). Let $1 \leq i < j \leq k$ be such that $P_j$ goes to the left of $C$ at their first common vertex $v$. The interior of $P_iP_j^{-1}$ is $\mathcal{K}_i^{ts} \cup \ldots \cup \mathcal{K}_{j-1}^{ts}$, which is disjoint from the interior of $C$. Then $P_i$ contains $v$, otherwise, by Lemma~\ref{lemma:two-cycles} applied to $P_iP_j^{-1}$ and $C$ at $v$, $P_j$ would go to the right of $C$ at $v$. Let $u$ be the first common vertex between $P_i$ and $C$.

If $u \neq v$ (Figure~\ref{fig:left-paths}, $(a)$), then as the interior of $D = (P_i)_{\bot{}v}(P_j)_{\bot{}v}^{-1}$ is disjoint from the interior of $C$, by applying Lemma~\ref{lemma:two-cycles} at $u$, $P_i$ goes to the left of $C$ at $u$.

If $u = v$ (Figure~\ref{fig:left-paths}, $(b)$), we apply Lemma~\ref{lemma:two-cycles} to the cycles $D, C$ and $D' = (P_i)_{v\top}(P_j)_{v\top}, C$ at $v$, we get that the following sequences of arcs appear in that order around $v$ (where $b^-_v$, $b^+_v$, $c_v^-$ and $c_v^+$ are arcs of $C$, occurring in that order around $v$):
\begin{itemize}
\item[$(i)$] $b_v^-, (P_j)_v^-, (P_i)_v^-, b_v^+$,
\item[$(ii)$] $c_v^-, (P_i)_v^+, (P_j)_v^+, c_v^+$ (then, because $P_i$ and $P_j$ do not cross $D$, $b^-_v = c_v^-$ and $b^+_v = c_v^+$),
\item[$(iii)$] and $c_v^-, P_j^+, P_j^-, c_v^+$ because $P_j$ goes to the left of $C$ at $v$.
\end{itemize}
Then, $P_i$ goes to the left of $C$ at $v$.

Thus, the paths that go to the left of $C$ at their first common vertices are of the form $P_1, \ldots, P_j$ for some $j$. The right case is symmetrical.
\end{proof}

We call a \emph{routing scheme} a function from $\mathcal{P}^2$ which gives the relative behaviour of each pair of paths at their first common vertex. A routing scheme is \emph{feasible} if there is an uncrossed solution respecting it. We denote $R := \max_{h \in E(H)} r(h) + 1$ and $k := |E(H)|$.

\begin{lemma}\label{lemma:routing-schemes}
For every network $(G,H,r,c)$ (with $G$ planar acyclic, $G+H$ Eulerian), there is a subset of at most $R^{4k^2}$ routing schemes that contains all the feasible routing schemes. Moreover, this subset can be polynomially enumerated.
\end{lemma} 

\begin{proof}
By Lemmas~\ref{lemma:crossing-intervals} and~\ref{lemma:non-crossing-intervals}, for each pair of demand arcs, it is sufficient to partition the paths for each demand arc into four intervals $A_1, B_1, C_1, D_1$ and $A_2, B_2, C_2, D_2$, where the relative behaviours of two paths are given by the following matrix :
\begin{equation*}
\begin{array}{c|cccc}
    & A_2 & B_2 & C_2 & D_2 \\
\hline
A_1 & C   & C   & L/L & L/R \\
B_1 & C   & C   & R/L & R/R \\
C_1 & L/L & L/R & C   & C \\
D_1 & R/L & R/R & C   & C \\
\end{array}
\end{equation*}
Here, $L/R$ means that the path $P_1$ for the first demand goes to the left of the path for the second $P_2$, while $P_2$ goes to the right of $P_1$. $C$ means they cross each other, and the others are defined similarly. Hence for each pair of demands, we have $R^8$ possible division in intervals, and as there are $\binom{k}{2}$ possible pairs, this gives an upper bound of $R^{4k(k-1)}$.
\end{proof}

From this, we derive an algorithm that tries every possible routing scheme. Given a routing scheme , we want to decide if it is feasible, and then to find an uncrossed solution. This is done by routing each vertex in increasing acyclic order, and making the paths grows from their origin. As $G+H$ is Eulerian and $G$ is acyclic, we have the property that in any solution, all the edges of the supply graphs are used by the solution. Thus, by routing in increasing acyclic order, when considering a vertex, every incoming arc is contained in a partial path already built. All we need to show is that, given a routing scheme, there is at most one way to grow the paths coming in the current vertex while respecting the routing scheme, which is the content of Lemma~\ref{lemma:unique-routing}.

\begin{proof} (of Theorem~\ref{theorem})

For each possible routing scheme determined by Lemma~\ref{lemma:routing-schemes}, we run the following algorithm: starting with empty paths, route every vertex by increasing order. By routing a vertex, we mean adding one arc to each path entering this vertex, while respecting the routing scheme, following Lemma~\ref{lemma:unique-routing}. For a given routing scheme, there are two possible failures:
\begin{itemize}
\item[-] it is not possible to route some vertex,
\item[-] the computed paths are not a solution, some demand is not satisfied.
\end{itemize}
The pseudo-polynomiality follows from the number of routing schemes, and the fact that routing a vertex $u$ can easily be done in polynomial time. First, we compute the relative behaviours, given either by the routing scheme, or by Lemma~\ref{lemma:second-intersection} (for that case, we need to precompute for each destination the set of vertices from which this destination is reachable), taking $O(d(u)^3)$. Then, we make the paths grow up in $O(d(u)^2)$. Note that $d(u) \leq Rk$, so the total complexity is $O(R^{4k^2 + 3} k^3 n)$.
\end{proof}

\section{Conclusion}

This result does not completely solve the planar acyclic arc-disjoint paths problem with the Eulerian condition, as the algorithm is only pseudo-polynomial. Finding an improvement to arbitrary requests would then be a next possible step. One could also try to extend this result to undirected graphs, or to directed graphs without the acyclicity condition on $G$, as both problems are open.

\bibliographystyle{plain-fr}
\bibliography{these}

\end{document}